\documentclass{iopart}%for normal type size
\usepackage{graphicx}

\usepackage{amssymb}
\usepackage{colordvi}
\usepackage{color}

\newcommand{\bq}{{\bf q}}
\newcommand{\ba}{{\bf a}}
\newcommand{\bb}{{\bf b}}
\newcommand{\bu}{{\bf u}}
\newcommand{\bv}{{\bf v}}

\newcommand{\bw}{{\bf w}}

\newcommand{\bS}{{\bf S}}
\newcommand{\bB}{{\bf B}}
\newcommand{\bE}{{\bf E}}
\newcommand{\tbw}{\tilde{\bf w}}

\newcommand{\tb}{\tilde{b}}
\newcommand{\tw}{\tilde{w}}

\newcommand{\vep}{\varepsilon }

\newcommand{\PT}{{\cal PT}}
\newcommand{\T}{{\cal T}}

\newcommand{\p}{{\cal P}}

\newcommand{\IM}{\textrm{Im$\,$}}
\newcommand{\RE}{\textrm{Re$\,$}}
\newcommand{\diag}{\mbox{diag}}

\newcommand{\lan}{\langle}
\newcommand{\ran}{\rangle}

\begin{document}

\title[Nonlinear lattices with $\PT$-symmetric linear part]{Stationary modes and integrals of motion in nonlinear lattices with $\PT$-symmetric linear part}

\author{D. A. Zezyulin$^{1}$ and V. V. Konotop$^{1,2}$ }
\address{
$^1$Centro de F\'isica Te\'orica e Computacional,   Faculdade de Ci\^encias, Universidade de Lisboa, Avenida Professor Gama Pinto 2, Lisboa 1649-003, Portugal
\\
$^2$ Departamento de F\'isica, Faculdade de Ci\^encias,
Universidade de Lisboa, Campo Grande, Ed. C8, Piso 6, Lisboa
1749-016, Portugal
}

\date{\today}

\begin{abstract}
We consider finite-dimensional nonlinear systems with linear part
described by a parity-time ($\PT$-) symmetric operator. We
investigate bifurcations of stationary   nonlinear modes  from the
eigenstates of the linear operator and  consider a class of
$\PT$-symmetric nonlinearities allowing for existence of the
families of  nonlinear modes. We pay particular attention to the
situations when the underlying linear $\PT$-symmetric operator is
characterized by the presence of degenerate eigenvalues or
exceptional-point singularity. In each of the cases we construct
formal expansions for small-amplitude nonlinear modes. We also
report a class of nonlinearities allowing for the  system to admit one or several integrals of motion, which turn out to be determined by the pseudo-Hermiticity of the  nonlinear operator.%
\end{abstract}

\submitto{\JPA}
%\maketitle

\section{Introduction}
\label{sec:back}

Richness of solutions is a typical feature of nonlinear problems.
Since a complete characterization of a vast set of all solutions
usually is not possible, a simpler problem of  classification of
admissible {\em types} of solutions of a nonlinear system becomes
a relevant and still nontrivial task. In this respect, there is an
important distinction between conservative and non-conservative
(i.e. dissipative) systems. Conservative nonlinear systems
typically possess continuous {\em families} of  nonlinear
stationary solutions (modes) which exist for fixed values of the
system parameters. A family of nonlinear modes can be parametrized
by an ``internal'' quantity (or few quantities), like, for
instance,
 $L^2$-norm of the mode  (which can correspond  to energy,  number
of particles, or total power, depending on the particular physical
statement)  or  its frequency (i.e. propagation constant or
chemical
potential). The situation becomes %completely
 different when one considers a
dissipative system characterized by the presence of gain and
losses. In order to admit a stationary solution, such a system
requires the absorbed energy to be compensated  exactly  by the
gain. The requirement of    energy balance imposes an additional
constraint on the shape of the solution which has to have a
nontrivial energy flow allowing for the energy transfer from the
gain regions to the lossy ones. As a result, for given values of
the system parameters, a typical dissipative nonlinear system
does not admit continuous families of nonlinear modes. Instead,
isolated nonlinear modes appear. From the dynamical  point of
view, these isolated modes typically behave as attractors (being
stable) or repellers (being unstable). In order to obtain a
continuous set of physically distinct dissipative nonlinear modes,
one has to vary parameters of the system, i.e. change the system
itself. Then, instead of continuous families, characteristic for
conservative systems with fixed parameters, the dissipative system
admits continuous {\em branches} of nonlinear modes which are
obtained through variation of the system parameters. We thus
distinguish between parametric families and branches. The
described dichotomy is particularly well-known in the context of
the discussion of dissipative solitons {\it vs.} conventional
solitons in the nonlinear Schr\"odinger-like and complex
Ginzburg--Landau equations~\cite{Akhmediev}.

In this context, the physical concept of parity-time ($\PT$)
symmetry~\cite{Bender} deserves a particular attention. This is
because the nonlinear  systems with $\PT$ symmetry  appear to
occupy an ``intermediate position'' between the conventional
conservative and dissipative systems. Whereas the  systems
governed by a  $\PT$-symmetric operators are of dissipative nature
and  require a solution to generate a nontrivial energy flow in
order to maintain itself, the exact gain and loss balance inherent
to the $\PT$ symmetry allows for the system  to possess in some
cases continuous families of nonlinear modes apart from the
branches typical to dissipative systems. There are numerous
studies where both branches and families of the nonlinear modes in
$\PT$-symmetric systems  have been reported. In particular, exact
solutions were reported for  the nonlinear $\PT$-symmetric dimer
(a system of two coupled nonlinear oscillators)
in~\cite{Ramezani}. Families of nonlinear discrete modes received
particular emphasis in \cite{ZK,birefring,KPT,vortexes},  while
branches of the solutions were  obtained for finite lattices like
$\PT$-symmetric  trimers and oligomers \cite{birefring,LK,LKMG},
as well as for  infinite $\PT$-symmetric chains \cite{KPT,
vortexes, MalomedRailway11, KPZ}.

Localized modes (or solitons) are also known to exist in extended
nonlinear systems, where they first were found in the presence of
a periodic $\PT$-symmetric potential  and cubic
nonlinearity~\cite{Musslimani}. Families of nonlinear modes in
continuous systems were   intensively investigated for gap
solitons in Kerr~\cite{Yang}  and $\chi^{2}$~\cite{Moreira} media,
in $\PT$-symmetric parabolic potential~\cite{ZK-parabolic}, in
defect $\PT$-symmetric lattices~\cite{defect}, in systems with
$\chi^{(2)}$ nonlinearity with embedded $\PT$-symmetric
defect~\cite{MAKY}, and    in the presence of a $\PT$-symmetric
superlattice~\cite{superlat}. Lattices with $\PT$-symmetric
\emph{nonlinear}  potentials  were proposed
in~\cite{nonlinearPT,nonlinearPT1}. Combined effect of linear and
nonlinear $\PT$-symmetric  lattices was explored in
discrete~\cite{Miroshnichenko11,Duanmu13} and
continuous~\cite{Micha} statements. Models of two coupled
waveguides (so-called $\PT$-symmetric nonlinear couplers) are also
known to support   solitons \cite{Driben11} and breathers
\cite{Barashenkov12}. Solitons in a  system consisting of a
necklace of optical  $\PT$-symmetric waveguides were recently
addressed in~\cite{Barashenkov13}.

In the most of the mentioned studies, the nonlinearity  inherent
to the system was fixed by the physical statement of the problem.
In optical applications one usually considers the  cubic (namely,
Kerr type) or quadratic ($\chi^{2}$-type) nonlinearities, while
realization of $\PT$ symmetry Bose-Einstein condensates
\cite{BEC3,BEC2,PTBEC} also implies  the Kerr type nonlinearity.
In many situations, however, the nonlinearity can be  changed,
which  makes it relevant to study physical systems obeying the
same linear properties  but having nonlinearities  of different
types.  \textcolor{black}{This leads us to the \emph{first~goal}
of the present  paper, which is the effect of the type of the
nonlinearity  on the existence and dynamics of nonlinear modes.}
In particular, we will study bifurcations of  families of
nonlinear modes from the eigenstates of the underlying linear
operator (i.e. the operator describing the linear part of the
system) and argue that the bifurcations are possible only if the
nonlinear operator admits a certain symmetry which can be
conveniently referred to as \textit{weak} $\PT$ symmetry. This
terminology is to emphasize that the class of weakly
$\PT$-symmetric nonlinear operators contains as a subset a class
of $\PT$-symmetric nonlinear operators  which (by analogy with the
definition of $\PT$ symmetry for linear operators) commute with
the $\PT$ operator. The \emph{second~goal} of our paper is to
study relevance of $\PT$-symmetric nonlinearities for dynamical
properties of the system. In particular, we will study
possibilities for the nonlinear system (\ref{dyn_main}) to admit
integrals of motion and show that if the nonlinearity is $\PT$
symmetric (and has a certain additional simple property), then the
nonlinear system does admit an integral of motion. On the other
hand, we will also argue that integrals of motion can  exist if
the nonlinear operator  is pseudo-Hermitian \cite{Mostaf2002}.

Apart from the type of nonlinearity,  properties of the nonlinear
modes may strongly depend on the character of the spectrum of the
underlying linear operator.  Most of the above mentioned studies
dealt with the situations when the linear  spectrum consists of
simple real eigenvalues. Recently, nonlinear modes bifurcating
from the doubly degenerate linear eigenstates  were reported
in~\cite{birefring}. \textcolor{black}{ The \emph{third~goal} of
the paper is to perform the analysis of bifurcations of stationary
nonlinear modes in a situation when the linear operator  has a
degenerate eigenstate of  finite multiplicity or
\textit{exceptional point} singularities~\cite{Kato}.}  For  both
those  cases  we develop formal asymptotical expansions which
describe bifurcations of the nonlinear modes from degenerate
linear eigenstates. In order to construct the expansions, we
explore the structure of the invariant subspace associated with
the multiple eigenvalue. For the case of a semi-simple eigenvalue,
we show that its invariant subspace can be spanned by a basis
consisting of $\PT$-invariant linearly independent eigenvectors
(see Proposition~1). For the situation when the geometric
multiplicity of the eigenvector is less than its algebraic
multiplicity, we show that there exists a $\PT$-invariant
generalized eigenvector (Proposition~2).

The organization of the paper is as follows. In
Sec.~\ref{sec:general}  we specify  the chosen model and make some
general remarks related to the  subject. Next, we address the
effect of the nonlinearity on bifurcation of nonlinear modes from
the eigenstates of the linear lattice. We consider the essentially
different situations of bifurcations from simple
(Sec.~\ref{sec:simple}) and  \textcolor{black}{semi-simple} %degenerate
(Sec.~\ref{sec:semisimple}) eigenvalues,   as well as the
nonlinear modes in  the presence of the  exceptional point
singularity (Sec.~\ref{sec:exc}). In Sec.~\ref{sec:intergral} we
address the question about the nonlinearities allowing  nonlinear
$\PT$-symmetric systems to have integrals of motion. The outcomes
are summarized in the Conclusion.

\section{The model and  general remarks}
\label{sec:general}

To proceed with our studies, we specify the chosen model.  In the
present work, we study  a  nonlinear system  of the form
\begin{eqnarray}
\label{dyn_main} i\dot{\bq} = - H(\gamma)\bq - F(\bq)\bq,
\end{eqnarray}
where $\bq=\bq(t)$ is a column-vector of $N$ elements, an overdot
stands for the derivative with respect to time $\dot\bq=d\bq/dt$
(or with respect to the propagation distance in  optical
terminology). The linear part  of the finite lattice
(\ref{dyn_main}) is described by a $N\times N$ symmetric matrix
$H(\gamma)$. We consider $\PT$-symmetric Hamiltonians $H(\gamma)$
\cite{Bender}, meaning that there exist parity, $\p$, and
time-reversal, $\T$, operators such that $\p^2=\T^2=I$, $[\p,
\T]=0$ and $[\PT,H]=0$ \textcolor{black}{(hereafter $I$ is the
identity operator). The mentioned  properties also imply that
$(\PT)^2 = I$.}

As it is customary, we define the time reversal operator $\T$ by
the complex conjugation, i.e. $\T\bq=\bq^*$ (hereafter an asterisk
stands for the complex conjugation), and consider  $\p$ to be a
matrix representation of the linear parity reversal operator. The
above definition of $\T$  ensures that   entries of the matrix
$\p$ are real~\cite{BMW}. Moreover,   $\PT$ symmetry implies
$H\p-\p H^*=0$, and hence
\begin{eqnarray}
\label{PT_sym} H^\dag = H^* = \p H \p,
\end{eqnarray}
i.e. the pseudo-Hermiticity of $H(\gamma)$ \cite{Mostaf2002}. We
also require the matrix  $\p$ to be  symmetric  (for the
discussion of the relevance of this requirement see \cite{BMW}):
\begin{equation}
\label{p} \p=\p^T=\p^\dag.
\end{equation}

The nonlinear operator $F(\bq)$ is a $N\times N$ matrix whose
elements depend on the field $\bq$.  We  focus
 on the case of the cubic nonlinearity when the  entries $F_{pj}(\bq)$ of the  matrix $F(\bq)$ are given as
\begin{equation}
\label{eq:F}
   F_{pj}(\bq)  =
   \bq^\dag {\cal F}_{pj}\bq =
   \sum_{l,m=1}^N
    %\sum_{q=1}^N
    {f}_{pj}^{lm}q_l^*  q_m, \quad p,j = 1, 2, \ldots, N,
\end{equation}
i.e. ${\cal F}_{pj}$ are  $N\times N$ matrices with
time-independent entries $f_{pj}^{lm}$, $l,m = 1, 2, \ldots, N$.
In other words,
  $F_{ij}(\bq)$ is a linear combination of    pair-wise products of the elements of the vectors $\bq(t)$ and $\bq^*(t)$ (some of the coefficients  ${f}_{pj}^{lm}$ can be equal to zero).
  \textcolor{black}{
Notice that if
${f}_{pj}^{lm}=\left({f}_{jp}^{ml}\right)^*={f}_{lj}^{pm}={f}_{pm}^{lj}$
then in the dissipationless limit [i.e. at values $\gamma$ for
which  $H(\gamma)$ becomes Hermitian] the system (\ref{dyn_main})
is Hamiltonian i.e.
 \begin{equation*}
 (F(\bq)\bq)_n= \frac{1}{2}\frac{\partial}{\partial q_n^*}\sum_{j,l,m,p=1}^N
       q_j^*q_l^* {f}_{jm}^{lp} q_mq_p, \quad n=1, 2, \ldots, N.
 \end{equation*}
The particular examples of nonlinearities $F(\bq)$ considered
below in this paper are of this type. Nevertheless, the analysis
we develop is applicable for a general case, which in particular
includes $\PT$-symmetric \emph{dissipative}
nonlinearities~\cite{nonlinearPT,nonlinearPT1,Miroshnichenko11,
Duanmu13}, when the  nonlinear operator $F(\bq)$  results in
losses and gain.}

Stationary nonlinear modes of system (\ref{dyn_main}) correspond
to solutions of the form $\bq(t)=e^{ibt}\bw$, where $\bw$ is a
time-independent column vector solving the stationary nonlinear
problem
\begin{equation}
\label{stationary}
    b\bw = H(\gamma)\bw + F(\bw)\bw, \qquad \mbox{Im}\,b=0.
\end{equation}
 Reality of $b$ is required for  the  existence of  stationary nonlinear modes,  and equality $F(\bw) = F(\bq)$ readily follows from
 Eq.~(\ref{eq:F}).  In the linear limit, which formally corresponds to $F(\bw)=0$, Eq.~(\ref{stationary}) is reduced to a linear eigenvalue  problem for the operator
$H(\gamma)$:
\begin{eqnarray}
\label{linear} \tb \tbw = H(\gamma)\tbw
\end{eqnarray}
(hereafter tildes stand to indicate the  eigenvectors and
eigenvalues belonging  to the linear spectrum). Let us now recall
some relevant properties of the linear problem~(\ref{linear})
(more details can be found e.g. in~\cite{Bender}). In a certain
range of values of the parameter $\gamma$, a non-Hermitian
$\PT$-symmetric operator $H=H(\gamma)$  may possess purely real
spectrum.  However, for a certain value (or values)  of $\gamma$
the system undergoes a spontaneous $\PT$ symmetry breaking,
corresponding to the transition from the real spectrum to complex
one (in the latter situation the system is said to be in the phase
of broken $\PT$ symmetry). The transition to the   broken $\PT$
symmetry phase  can be described in terms of
\emph{exceptional-point} spectral
singularity~\cite{Graefe11,Ramezani12}.
%
%If all the eigenvalues $\tb$ are real we say  it is said that the $\PT$-symmetry is unbroken %(otherwise the  $\PT$-symmetry is broken).

If an eigenvalue $\tb$ of  $H(\gamma)$ is real and possesses
exactly one linearly independent eigenvector   $\tbw$, then the
latter can be chosen to be \emph{$\PT$-invariant}~\cite{BBJ}, i.e.
\begin{eqnarray}
\label{PTw} \PT\tbw=\tbw,
\end{eqnarray}
independently on whether the $\PT$ symmetry of $H(\gamma)$ is
unbroken or
broken. %The following note is in order here.
Notice that   condition (\ref{PTw}) fixes the phase of the vector
$\tbw$, while the model (\ref{dyn_main}) is phase invariant thanks
to the choice of the cubic nonlinearity of the form (\ref{eq:F}).
Instead of vector $\tbw$ fixed by (\ref{PTw}), one can consider
any eigenvector of the form    $e^{i\varphi}\tbw$, where $\varphi$
is real. However, due to the phase invariance, this generalization
does not lead to physically distinct solutions. Therefore, to
simplify the algebra below we hold the definition  of the
$\PT$-invariant mode (\ref{PTw}),
 bearing in mind that modes (either linear or nonlinear) $\bw$ and  $e^{i\varphi}\bw$ are physically equivalent.

Equation~(\ref{PTw}) trivially leads to the properties $
\tbw^*=\T\tbw=\p\tbw$. Introducing the inner product as $ \lan
{\bf a},{\bf b}\ran = {\bf a}^\dag{\bf b}=\sum_{j=1}^N a_j^*b_j $
we arrive at the conclusion that  $\langle \tbw^*, \tbw\rangle$ is
real. Indeed, for any two  $\PT$-invariant vectors $\ba=\PT \ba$
and $\bb=\PT\bb$ one verifies that
 \begin{eqnarray}
\label{prod1} \langle \ba^*, \bb\rangle=\langle \p\ba, \bb\rangle
= \langle \p\ba, \PT\bb\rangle =   \langle \ba, \T\bb\rangle
%= \langle \ba, \p\bb\rangle
= \langle \ba, \bb^*\rangle=\langle \ba^*, \bb\rangle^*.
 \end{eqnarray}

In the situation when the eigenvalue $\tb$ is multiple, one should
distinguish between two different cases. In the first case, the
multiple eigenvalue $\tb$ (with   algebraic multiplicity equal to
$n$) possesses an invariant subspace spanned by $n$ linearly
independent eigenvectors. In other words, the eigenvalue $\tb$ is
\textit{semi-simple}, and the  Hamiltonian  $H(\gamma)$  is
diagonalizable (provided that all other eigenvalues are also
simple or semi-simple). This situation arises, in particular, when
two eigenvalues of a  parameter-dependent Hamiltonian $H(\gamma)$
coalesce at some value of the control parameter $\gamma$ but the
two corresponding eigenvalues remain linearly independent.
Physically such situation appears, in particular, when two (or
more) identical systems are linearly coupled: the degeneracy
occurs when the coupling becomes zero.

The second case   corresponds to the situation when collision of
the eigenvalues is accompanied by collisions of eigenvectors. Then
the geometrical multiplicity of the eigenvalue $\tb$ if less than
the algebraic multiplicity, i.e. the dimension of the invariant
subspace of the  multiple eigenvalue $\tb$ is less than $n$. This
corresponds to so-called {\em exceptional} points~\cite{Kato}. As
it was mentioned above, the relevance of such points is in
particular related to the transition to the phase of broken $\PT$
symmetry (see also the discussion in \cite{Heiss}).

\section{Bifurcations of the families of   nonlinear modes. Simple eigenvalue}
\label{sec:simple}

%It has been found in~\cite{ZK}, that existence of the continuous
%families of nonlinear modes in systems described by a
%$\PT$-symmetric operator $H(\gamma)$ essentially depends on the
%specific form of the nonlinearity $F(\bq)$ (for a discussion of
%relevance of the nonlinear properties with respect to the $\PT$
%symmetry see also~\cite{LKMG}). In particular, it was found that
%families of nonlinear modes exist only  for a specific relation
%between the symmetries of the linear and nonlinear parts of the
%system. In other words, existence of continuous families of
%nonlinear modes is  not a general  property automatically
%guaranteed by $\PT$ symmetry of the linear part of the system
%(\ref{dyn_main}).  The analysis performed in \cite{ZK} was based
%on the study of conditions for the underlying  linear
%$\PT$-symmetric system to allow for bifurcations of the families
%of nonlinear modes from the linear eigenstates of $H(\gamma)$.  It
%has also been proposed in~\cite{KPZ} that the existence of the
%families of nonlinear modes can be established through the
%analytical continuation from the the anticontinuum limit (the
%technique well known in the theory of conservative
%lattices~\cite{MA}). Bifurcations of the nonlinear modes from the
%eigenstates of the underlying  linear lattices,
%%
%as well as from the anticontinuum limit, were recently addressed
%in a rigorous mathematical framework~\cite{KPT}.

We start by considering families of nonlinear modes  bifurcating
from a linear eigenstate of the $\PT$-symmetric   Hamiltonian $H$.
Let $\tb$ be a simple real eigenvalue and $\tbw$ is the
corresponding eigenvector solving linear problem (\ref{linear}).
According to the discussion in Sec.~\ref{sec:general}, without
loss of generality we can assume that $\tbw$ is $\PT$ invariant,
i.e. satisfies the condition  (\ref{PTw}).

In the vicinity of the \textit{linear limit} the nonlinear modes
bifurcating from the eigenstate $\tbw$  can be described using
formal expansions~\cite{ZK}
\begin{eqnarray}
\label{expan1} \bw = \varepsilon\tbw +\vep^3\bw^{(3)} +
o(\varepsilon^3) \quad \mbox{and}  \quad b = \tb +
\varepsilon^2b^{(2)} + o(\varepsilon^2),
\end{eqnarray}
where $\varepsilon$ is a real small parameter, $\varepsilon\ll 1$,
and without loss of generality we impose the normalization
condition $\langle\tbw,\tbw\rangle=1$. Coefficients $\bw^{(3)}$
and $b^{(2)}$ of the expansions are to be determined.

Substituting expansions (\ref{expan1}) into Eq.~(\ref{stationary})
and noticing from Eq.~(\ref{eq:F}) that $F(\bw) = \vep^2F(\tbw) +
O(\vep^3)$, one arrives at the following equation for the shift of
the eigenvalue $b^{(2)}$:
\begin{equation}
\label{second_1}
b^{(2)}\tbw  = (H-\tb)\bw^{(3)} + F(\tbw)\tbw.
\end{equation}
Multiplying this equation by  $\tbw^*$ and using  %the relations
%\begin{eqnarray}
%\langle \tbw^*,H\tbw\rangle=\langle \p H\PT\tbw,\tbw\rangle=\langle \T H\tbw,\tbw\rangle=\tb\langle \tbw^*,\tbw\rangle
%\end{eqnarray}
%valid due to
Eqs.~(\ref{PT_sym}) and (\ref{linear}),
  one readily obtains
\begin{equation}
\label{eq_p} b^{(2)} =  \frac{\langle\tbw^*, { F}(\tbw)\tbw
\rangle }{{\langle \tbw^*, \tbw\rangle}}.
\end{equation}
Taking into account   reality of the  product $\langle \tbw^*,
\tbw\rangle$ [see  Eqs.~(\ref{prod1})], one notices that
bifurcation of a family nonlinear modes is possible only if
\begin{eqnarray}
\label{cond2}
 \IM \langle \tbw^*, {F}(\tbw)\tbw \rangle = 0,
 %= \IM \langle\C\tbw_n, {\bf F}(\tbw_n)\tbw_n \rangle =0
\end{eqnarray}
which is necessary for  $b^{(2)}$  to be  real.
Thus (\ref{cond2}) is a necessary condition % (\ref{cond1}) and
   for a family of nonlinear modes to bifurcate from the nondegenerate eigenstate of $H$ corresponding to $\tb$.

By analogy with the condition (\ref{PTw}) which guarantees reality
of the denominator of the right hand side of Eq.~(\ref{eq_p}), in
order to ensure that the numerator is also real,  we introduce the
following condition for the nonlinear operator $F(\bw)$ \cite{ZK}: %One can readily establish a class of the nonlinearities which satisfy the both theses %conditions. These are the nonlinearities such that
 \begin{equation}
 \label{eq-NL1}
\PT{ F}(\bw)\bw = { F}(\bw)\bw \mbox{\quad for all \quad} \bw
\quad \mbox{such that \quad }  \PT\bw = \bw.
%\mbox{\quad and  \quad} \C{\bf F}(\bw)\bw = {\bf F}(\bw)\bw
\end{equation}
 %for all $\bw$ such that  $\PT\bw = \bw$.
Obviously, this condition is equivalent to $[\PT,{ F}(\bw)]=0$, where the latter commutator is only considered on the set  of vectors satisfying  $\PT\bw = \bw$.   In what follows the nonlinearities obeying~(\ref{eq-NL1}) will be
%referred to as $\PT$-symmetric and
said to be  {\it weakly} $\PT$ {\it symmetric}. The set of weakly
$\PT$-symmetric nonlinearities  corresponding to the given parity
operator $\p$  will be denoted by $NL_{w\PT}(\p)$.  Thus the weak
$\PT$ symmetry of the nonlinear operator $F(\bw)$ appears to be a
necessary  condition for the system (\ref{dyn_main}) to admit
continuous families of nonlinear modes.
 %(notice that in~\cite{LKMG} the nonlinearity satisfying property (\ref{eq-NL1}) was termed  as  simply $\PT$ symmetric).

The choice of the term \emph{weak} $\PT$ symmetry
%  of the
can be understood if one considers a more restrictive condition
for the nonlinear operator $F(\bw)$ to commute with the  $\PT$
operator for \emph{any} vector $\bw$:
\begin{eqnarray}
\label{GPTnonlin} [\PT,{ F}({\bw})]=0.
\end{eqnarray}
Nonlinear operators $F(\bw)$ of the latter type will be said to be
$\PT$ symmetric.   The set of $\PT$-symmetric nonlinearities
corresponding to the given parity operator $\p$   will be
 denoted as $NL_{\PT}(\p)$.  Obviously,   for a given parity operator $\p$, the class of   weakly  $\PT$-symmetric nonlinearities  contains the class of  $\PT$ symmetric nonlinearities as a subset:   $NL_{\PT}(\p) \subset NL_{w\PT}(\p)$.
The relevance of the $NL_{\PT}$-type nonlinearities will be
discussed in Sec.~\ref{sec:intergral}.

%Notice that due to possibility of existing of several $\p$
%operators, in the brackets we indicate the symmetry with respect
%which parity operators is considered.

As an example, let us consider a finite nonlinear  system with  $N=4$ (i.e. a quadrimer) and
%
%$4\times 4$ parity operators  $\p$, which can be expressed in terms of the Pauli matrices %(see also Ref.~\cite{LKMG}): % following , we denote by
%\begin{eqnarray}
%\label{p}
%\p_{ij}=\sigma_i\otimes \sigma_j
%\end{eqnarray}
% %Indeed if (\ref{GPTnonlin}) is satisfied, then $\PT{ F}(\bw)\bw={ F}(\bw)\bw={ %F}(\bw)\PT\bw.$
%As an  example, we mention
a Kerr  nonlinearity, $F_{K}(\bw)$, whose elements are defined by  $F_{K,pj}(\bw)= \delta_{pj}|w_j|^2$ where
 $\delta_{pj}$ is the Kronecker delta, and $p,j=1,..,4$.  This nonlinearity satisfies the requirement (\ref{eq-NL1}) provided that the parity operator is chosen as
\begin{equation}
\label{eq:P11}
\p_{11}  = \sigma_1\otimes \sigma_1=\left (\!\begin {array}{cccc}
0 & 0 & 0 & 1\\%
0&0&1&0\\%
0&1 &0& 0\\%
1&0 &0&0\end {array} \! \right)
\end{equation}
(hereafter $\sigma_j$ are the Pauli matrices). However, the
nonlinearity  $F_{K}(\bw)$   does not satisfy the condition
(\ref{GPTnonlin}). Therefore $F_{K}\in NL_{w\PT}(\p_{11})
\setminus NL_{\PT}(\p_{11})$.

Notice that expansions (\ref{expan1}) suggest that the nonlinear
modes $\bw$ bifurcating from the  $\PT$-invariant  eigenvector
$\tbw$ will also  be $\PT$ invariant: $\PT\bw=\bw$. This fact is
confirmed in numerous  previous studies, see e.g.~\cite{ZK,
KPT,vortexes, LKMG, KPZ, Yang,BEC3}, where  a large library of
results dedicated to  nonlinear modes bifurcating from the
non-degenerate (simple) linear eigenstates and supported by the
nonlinearities of the $NL_{w\PT}$ type can be found. Existence of
the families of nonlinear modes was also established through the
analytical continuation from the the anticontinuum limit
\cite{KPZ} (the technique well known in the theory of conservative
lattices~\cite{MA}). Bifurcations of the nonlinear modes from the
simple eigenvalues   of the underlying  linear lattices, as well
as from the anticontinuum limit, were recently addressed in a
rigorous mathematical framework~\cite{KPT}.

\section{Bifurcation from the semi-simple eigenvalue}
\label{sec:semisimple}

\subsection{Structure of the invariant subspace}

In the previous section  we have considered bifurcations  of
nonlinear modes from a nondegenerate (simple) eigenvalue of the
underlying linear problem. Now we turn to the case of  a
  semi-simple eigenvalue, i.e. a multiple eigenvalue
whose geometric and algebraic multiplicities are equal.
%In this section,  we are considering
Let  $\tb$ be a real semi-simple eigenvalue   of   multiplicity
equal to $n$.  This means that there exist $n$ linearly
independent eigenvectors $\tbw_j$, $j=1,\ldots, n$, such that
$H\tbw_j = \tb \tbw_j$, and there is a $n$-dimensional
$H$-invariant subspace spanned by $\tbw_j$ such that any vector
from this subspace is an eigenvector of the operator $H$.

Adopting the approach of formal expansions used in
Sec.~\ref{sec:simple}  for the case of  a simple eigenvalue, we are
going to construct a family of %\PT$-invariant
nonlinear modes which in the vicinity of the bifurcation from the
linear limit behave as $\bw = \vep\tbw+ \ldots$,
\textcolor{black}{where $\tbw$ is some eigenvector of $H$
corresponding to the eigenvalue  $\tb$. Looking for
$\PT$-invariant nonlinear modes $\PT\bw=\bw$, one has to find a
$\PT$-invariant eigenvector   $\tbw$  for the expansion to be
valid. However, }
%
%
%, where $\tbw$ is a $\PT$-invariant
%eigenvector of $H$ corresponding to $\tb$.
while in the case of a
simple eigenvalue a $\PT$-invariant eigenvector  $\tbw$
necessarily exists (and is unique up to irrelevant multiplier), in
the case of a  semi-simple  eigenvalue $\tb$ it is not readily
obvious how many $\PT$-invariant eigenvectors exist  (if any).
 Therefore, in the situation at hand it is necessary to explore structure of the invariant subspace associated with   the multiple eigenvalue $\tb$.
%As we will
%
%
Let us show that in the invariant subspace of $\tb$ one can always
find a basis of $n$ linearly independent $\PT$-invariant
eigenvectors (independently on whether the $\PT$ symmetry of $H$
is broken or not).

%
%
%To this end,  let us first prove the following
%simple but useful proposition which shows that the whole linear
%space has a complete basis of $\PT$-invariant vectors.
%
%\medskip
%\textbf{Proposition 1.} Let a parity  operator  $\p$ acts in an
%$N$-dimensional linear space.  Then there exists a complete basis
%$(\be_1, \be_2, \ldots, \be_N)$ consisting of $N$ $\PT$-invariant
%vectors $\PT\be_j=\be_j$, $j=1, 2, \ldots, N$.
%
%\textbf{Proof.}  The most general $N\times N$ matrix $\p$ can be expressed as $\p = R \p_0 R^{-1}$, where $\p_0$ is a diagonal matrix having on the diagonal $m_+$  entries equal to 1 and $m_-=N - m_+$ entries equal to $-1$, and $R$ is an arbitrary orthogonal matrix \cite{BMW}. Columns of $R$ are the eigenvectors of $\p$ and they compose a complete basis of real-valued vectors (all the entries of an orthogonal matrix are real). Then a $\PT$-invariant basis can be found using to the following rule: if $j$-th entry on the main diagonal of $\p_0$ is  $1$ then $\be_j$ is taken as $j$-th column of $R$. If the $j$th entry is equal to $-1$, then $\be_j$     is taken as $j$-th column of $R$ multiplied by  the imaginary unit $i$. Then the condition  $\PT\be_j=\be_j$  can be verified straightforwardly.~\hfill $\blacksquare$
%\medskip

\medskip
\textbf{Proposition 1}.  Let $\tb$ be a real semi-simple
eigenvalue  of  multiplicity $n$. Then the invariant subspace of
$\tb$ has a complete basis $(\bu_1, \bu_2, \ldots, \bu_n)$ of
$\PT$-invariant eigenvectors: $\PT \bu_j= \bu_j$, $j=1, 2, \ldots,
n$.

%\textbf{Remark}. For $n=1$ the statement of Proposition~2 is nearly trivial  since in this case eigenvalue $\tb$ becomes simple and the situation is described in Sec.~\ref{sec:simple}. For $n=N$ the statement of Proposition~2 is also obvious because in this case the invariant subspace coincides with the whole $N$-dimensional linear space and one can set $\bu_j=\be_j$, where $\be_j$ are given by  Proposition~1.

\smallskip

\textbf{Proof of Proposition~1.} The condition  of the proposition
implies that there exist $n$ linearly independent eigenvectors
$\tbw_j$, $j=1, 2, \ldots, n$, such that $H\tbw_{j}=\tb \tbw_{j}$.
If the choice  of the linearly independent eigenvectors  $\tbw_j$
is arbitrary, then generally speaking $\PT\tbw_j \ne \tbw_j$, i.e.
eigenvectors $\tbw_j$ (or some of them) are not  $\PT$ invariant.
Thus, in order to prove the proposition we must find $n$ linearly
independent $\PT$-invariant eigenvectors.

%To this end, let us represent the eigenvectors  $\tbw_j$  in terms
%of a $\PT$-invariant basis $(\be_1, \be_2, \ldots, \be_N)$ which
%exists due to Proposition~1: $\tbw_{j}=\sum_{k=1}^NC_{jk}\be_k$
%($j=1,..,n$). The coefficients $C_{jk}$ are considered as entries
%of the $n\times N$ matrix $C$ of rank $n$.
To this end we apply $\PT$ operator to each $\tbw_j$ and obtain a
new set of vectors $\bv_j = \PT \tbw_j$, $j=1,\ldots, n$. $\PT$
symmetry of $H$ and reality of $\tb$ imply that $H \bv_j = \tb
\bv_j$, i.e. each  $\bv_j$ belongs to the invariant subspace of
the eigenvalue $\tb$. \textcolor{black}{ Notice also that   $\PT
\bv_j = \bw_j$ [thanks to $(\PT)^2=I$]. Then linear independence
of vectors $\tbw_j$ implies that the vectors $\bv_j$ are also
linearly independent and therefore constitute a basis in the
invariant subspace of  $\tb$. Therefore, there exists a  $n\times
n$ nonsingular matrix $D$  such that $\bv_j =
\sum_{k=1}^{n}D_{jk}\tbw_k$.}

\textcolor{black}{ Let us now introduce a new set of vectors
$\bu_j$, $j=1,2, \ldots, n$, given as
\begin{eqnarray*}
\hspace{-1cm} \bu_j = e^{i\phi}\tbw_j + e^{-i\phi}\bv_j= e^{-i\phi}\left(\sum_{k=0}^n D_{jk}\tbw_k + e^{2i\phi}\tbw_j\right) = e^{-i\phi} \sum_{k=0}^n M_{jk} \tbw_k,
\end{eqnarray*}
where $\phi$ is an arbitrary (so far) real parameter, and
$M_{jk}$ are  the entries of the $n\times n$ matrix $M$ defined as
$M=D+e^{2i\phi}I$  (here $I$ is the $n\times n$ identity matrix).
Then each $\bu_j$    belongs to the invariant subspace of $\tb$ and is obviously $\PT$ invariant. % thanks to  reality of  the matrix  $(e^{-i\phi}C^* + e^{i\phi}C)$.  Obviously
Besides, one can always find such $\phi$ that  matrix  $M$  is
nonsingular  (it is necessary and  sufficient to choose $\phi$
such that $\rho=-e^{2i\phi}$  does not belong to the spectrum  of
$D$). Once $M$ is nonsingular, the  eigenvectors  $\bu_j$ are
linearly independent and therefore constitute a complete
$\PT$-invariant basis  in the invariant subspace associated with
the eigenvalue $\tb$. Thus the proposition is proven.~\hfill
$\blacksquare$}

\medskip

\subsection{Expansions for nonlinear modes}

Turning now to bifurcations of nonlinear modes, we employ  the
same expansions as in the case of a simple eigenvalue:
\begin{eqnarray}
\label{expan3}
\bw = \varepsilon\tbw
%+\vep^2\bw^{(2)}
+ \vep^3\bw^{(3)} + o(\varepsilon^3), \quad \mbox{and}  \quad b
= \tb + \varepsilon^2b^{(2)} + o(\varepsilon^2).
\end{eqnarray}
%as for the case of nongedenerate eigenvalue, i.e. (\ref{expan1}), where
However, now the vector $\tbw$ is a linear combination of the  $\PT$-invariant eigenvectors  $\bu_j$ constructed in Proposition~1: $\tbw = \sum_{j=1}^n c_j \bu_j$. In order to assure that the linear solution $\tbw$ is $\PT$-invariant, i.e. satisfies (\ref{PTw}), we require all the coefficients $c_j$ to be real. %It turns out
However,
 an arbitrary set  of  coefficients $c_j$ generally speaking
 %
 %that not arbitrary set of $c_j$ yields
 does not represent a linear mode $\tbw$ allowing for a bifurcation of a family of nonlinear modes $\bw$. The relevant relations among the coefficients can be  found by multiplying Eq.~(\ref{second_1}) by
$\bu_j^*$
 leading to conditions as follows
%\begin{equation}
%\label{set_c}
%\mbox{for each $j=1,2, \ldots n$:\quad} b^{(2)} =  \frac{\langle \bu_j^*, {F}(\tbw)\tbw  \rangle }{{\langle \bu_j^*, %\tbw\rangle}},
%\end{equation}
\begin{equation}
\label{set_c}
\mbox{for each $j=1,2, \ldots n$:\quad} b^{(2)} =  \frac{\langle \bu_j^*, {F}(\tbw)\tbw  \rangle }{{\langle \bu_j^*, \tbw\rangle}},
\end{equation}
where $b^{(2)}$ is required to be real. One can readily ensure that if the nonlinearity ${F}(\bw)$  satisfies to requirement (\ref{eq-NL1}), i.e. ${F}(\bw)\in NL_{w\PT}(\p)$, then $b^{(2)}$ is real. Indeed:
\begin{eqnarray*}
\frac{\langle\bu_j^*, {F}(\tbw)\tbw  \rangle^*}{{\langle \bu_j^*, \tbw\rangle}^*}=\frac{\langle \PT\bu_j, \T({F}(\tbw)\tbw)  \rangle }{{\langle \PT\bu_j, \T\tbw\rangle}}%=
%\frac{\langle \bu_j, \T\PT{F}(\tbw)\tbw  \rangle }{{\langle \bu_j, \T\PT\tbw\rangle}}
%\\
= \frac{\langle  \bu_j^*, {F}(\tbw)\tbw  \rangle }{{\langle \bu_j^*, \tbw\rangle}}.
\end{eqnarray*}

Bearing in mind the normalization condition $\langle \tbw, \tbw\rangle = 1$,
the system~(\ref{set_c}) can be viewed as $n$ algebraic equations with respect to $n-1$ independent real coefficients $c_j$; the requirement of   %solvability
 compatibility of these equation determines $b^{(2)}$.  The existence and a number of the solutions obviously depends on the specific form of the nonlinearity $F(\bw)$, the latter determining  the diversity of physically distinct families of nonlinear modes bifurcating from  the eigenstates of the linear spectrum.

Finally, we mention a peculiar but physically relevant situation
when for a certain set of the coefficients $c_j$ the eigenstate
$\tbw$  becomes   an eigenvector of the nonlinear eigenvalue
problem  $F(\tbw)\tbw=\lambda \tbw$, and the  eigenvalue $\lambda$
is real. Then Eqs.~(\ref{set_c}) are automatically satisfied
provided that $b^{(2)}=\lambda$, i.e. $b^{(2)}=\langle \tbw,
F(\tbw)\tbw \rangle$. In this case the number of free parameters
exceeds the number of the imposed constraints. Such a situation
has been encountered for example  in the model of a
$\PT$-symmetric birefringent coupler~\cite{birefring} and occurs
in the example considered in the next subsection.

\subsection{Example: $\PT$-symmetric coupler of bi-chromatic light}
\label{sec:polychrom}

As we already mentioned, the presence of multiply degenerated
eigenvalues in the linear spectrum typically occurs when two
identical systems are decoupled linearly, and coupling exists only
due to the nonlinearity. Let us consider the following  example:
\numparts
\begin{eqnarray}
\label{ex-1} i\dot{q}_1=i\gamma q_1+ q_2+(|q_1|^2 +|q_3|^2)q_1,
\\
\label{ex-2} i\dot{q}_2=-i\gamma q_2+ q_1+\varkappa (|q_2|^2
+|q_4|^2)q_2,
\\
\label{ex-3} i\dot{q}_3=-i\gamma q_3+ q_4+(|q_3|^2 +|q_1|^2)q_3,
\\
\label{ex-4} i\dot{q}_4=i\gamma q_4+ q_3+\varkappa (|q_4|^2
+|q_2|^2)q_4.
\end{eqnarray}
\endnumparts
It can be viewed as a  model for  propagation of a bi-chromatic light in a $\PT$-symmetric coupler. Then $q_j$ and $q_{j+2}$ ($j=1,2$) are the fields having the propagation constants $k_1$ and $k_2$ in $j$-th waveguide, where self-phase and cross-phase modulation of the two modes propagating in one arm are normalized to one. The modes $k_1$ and $k_2$ are subject to gain and absorption in the first waveguide (and {\it vice versa} in the second waveguide). The real parameter $\varkappa$ describes the difference in the Kerr nonlinearities in the waveguides.

The linear Hamiltonian of (\ref{ex-1})-(\ref{ex-4}) is described by the matrix $H_{1}(\gamma,-\gamma)$ where
 \begin{equation}
 \label{Hbc}
H_{bc} (\gamma_1,\gamma_2)= \left(\begin {array}{cccc}
i\gamma_1 & 1 & 0 & 0\\%
1&i\gamma_2&0&0\\%
0&0 &-i\gamma_1& 1\\%
0&0 &1&-i\gamma_2\end {array} \right)
\end{equation}
and the nonlinearity is given by
\begin{eqnarray}
\label{F_polychrom}%
 F_{bc}(\bq) = \left(\!\!\begin {array}{cccc}
\! |q_1|^2 +|q_3|^2   & 0 & 0 & 0\\%
0&
\!\!\varkappa(|q_2|^2 +|q_4|^2)  &0&0\\%
0&0 &|q_1|^2 +|q_3|^2 & 0\\%
0&0 &0&\!\!\varkappa(|q_2|^2 +|q_4|^2)  \end {array}\!\! \right).
%\nonumber \\
%   F=\mbox{diag}(f_{1}|q_1|^2 +f|q_3|^2,f_{1}|q_2|^2 +f|q_4|^2,f_{2}|q_3|^2 %+f|q_1|^2,f_{2}|q_4|^2 +f|q_2|^2)
\nonumber \\
\end{eqnarray}
Subscripts ``$bc$'' in Eqs.~(\ref{Hbc}) and (\ref{F_polychrom})
stand for ``\textit{bi-chromatic}''.

\begin{figure}
\centerline{
\includegraphics[width=1.0\textwidth]{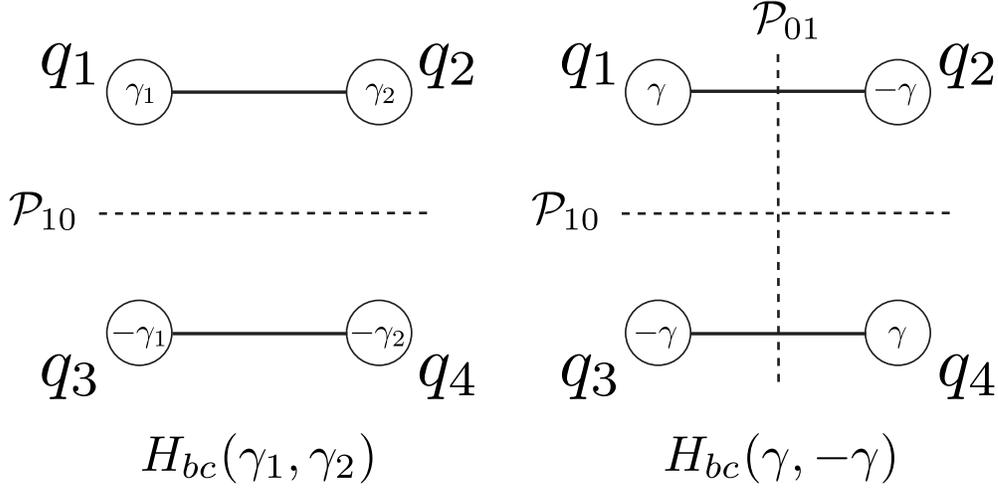}
} \caption{The graphs of the Hamiltonians
$H_{bc}(\gamma_1,\gamma_2)$ (left graph) and
$H_{bc}(\gamma,-\gamma)$ (right graph) and the symmetries
corresponding to the  respective parity operators.
%(notice that the for identifying the position of the symmetry axes
%one should coincide the absolute values of the gain/loss
%coefficients $\gamma$ since the $\T$ operator is the complex
%conjugation.
}
 \label{graph}
\end{figure}

One readily verifies that $H_{bc}(\gamma_1, \gamma_2)$ is
$\p_{10}\T$-symmetric with respect to
\begin{eqnarray}
\label{eq:Ps}
 \p_{10}=\sigma_1\otimes \sigma_0= \left(\!\begin {array}{cccc}
0 & 0 & 1 & 0\\%
0&0&0&1\\%
1&0 &0& 0\\%
0&1 &0&0\end {array} \!\right).
\end{eqnarray}
In the particular case $\gamma_1=-\gamma_2 = \gamma$
\textcolor{black}{an additional symmetry appears. Indeed,}
$H_{bc}(\gamma, -\gamma)$ is also  $\p_{01}\T$-symmetric with
\begin{eqnarray}
\p_{01}=\sigma_0\otimes \sigma_1= \left(\!\begin {array}{cccc}
0 & 1 & 0 & 0\\%
1&0&0&0\\%
0&0 &0& 1\\%
0&0 &1&0\end {array} \!\right).
\nonumber
\end{eqnarray}
\textcolor{black}{The physical meaning of the operators $\p_{10}$
and $\p_{01}$ becomes particularly clear in terms of the graph
representation~\cite{ZK}, as this is illustrated in
Fig.~\ref{graph}. }

One also  verifies that
%\begin{eqnarray}
$F_{bc}(\bq) \in NL_{\PT}(\p_{10}) \subset  NL_{w\PT}(\p_{10})$.
%\end{eqnarray}
At $\varkappa=1$ one additionally has $F_{bc}(\bq) \in
NL_{wPT}(\p_{01})$.

As it is clear, the operator  $H_{bc}(\gamma,-\gamma)$  represents
two uncoupled linear $\p\T$-symmetric dimers.  Spectrum of
$H_{bc}(\gamma,-\gamma)$ consists of  two     double semi-simple
eigenvalues  given by $\tb_\pm=\pm \sqrt{1-\gamma^2}$. The
$\p_{10}\T$-invariant eigenvectors of $H_{bc}(\gamma,-\gamma)$
(which exist due to Proposition~1) can be written does  as
follows:
\begin{eqnarray*}
 \bu_{+,1}=\left(\!\!
 \begin{array}{c}
 e^{i\varphi} \\ e^{- i\varphi}\\ e^{-i\varphi} \\ e^{i\varphi}
 \end{array}
 \!\!\right),
 \,\,\,\,\,\,
 \bu_{+,2}=i\left(\!\!
 \begin{array}{c}
 e^{i\varphi} \\ e^{- i\varphi}\\ -e^{-i\varphi} \\ -e^{i\varphi}
 \end{array}
 \!\!\right),%
  \\[2mm]
  \bu_{-,1}=\left(\!\!
 \begin{array}{c}
 -e^{-i\varphi} \\ e^{i\varphi}\\ -e^{i\varphi} \\ e^{-i\varphi}
 \end{array}
 \!\!\right),
 \,\,\,
 \bu_{-,2}=i\left(\!\!
 \begin{array}{c}
 -e^{-i\varphi} \\ e^{ i\varphi}\\ e^{i\varphi} \\ -e^{-i\varphi}
 \end{array}
 \!\!\right),
\end{eqnarray*}
where $\varphi$ is defined by requirements $\sin(2\varphi)=\gamma$ and $\cos(2\varphi)=\sqrt{1-\gamma^2}$.

Let us search for nonlinear modes bifurcating from the eigenvalue
$\tb_+$ (analysis for the eigenvalue $\tb_-$ yields   similar
results). Following to the above approach, we  search for the
linear eigenvector in the form as a linear combination $\tbw = c_1
\bu_{+, 1} + c_2 \bu_{+,2}$ where $c_{1,2}$ are real coefficients.
Subject to the normalization   $\langle \tbw,  \tbw \rangle = 1$,
we obtain $c_1^2 + c_2^2 = 1/4$. \textcolor{black}{Straightforward
algebra gives %
\numparts %
\begin{eqnarray}
\label{eq:uFw1}%
\langle \bu_{+, 1}^*, F_{bc}(\tbw )\tbw \rangle = c_1\cos(2\varphi)(\varkappa+1) + c_2\sin(2\varphi)(\varkappa-1),\\[1mm]
\langle \bu_{+, 2}^*, F_{bc}(\tbw )\tbw \rangle = c_1\sin(2\varphi)(\varkappa-1) - c_2\cos(2\varphi)(\varkappa+1),\\[1mm]
\label{eq:uFw3} \langle \bu_{+, 1}^*, \tbw \rangle =
4c_1\cos(2\varphi), \quad \langle \bu_{+, 2}^*, \tbw \rangle =
-4c_2\cos(2\varphi).
\end{eqnarray}
\endnumparts
Next, following Eqs.~(\ref{set_c}), we require %
\numparts \label{eq:bichr0}
\begin{eqnarray}
\label{eq:bichr1}%
b^{(2)}\langle \bu_{+, 1}^*, \tbw \rangle = \langle \bu_{+, 1}^*, F_{bc}(\tbw )\tbw \rangle,\\%[0.5mm]
\label{eq:bichr2}%
b^{(2)}\langle \bu_{+, 2}^*, \tbw \rangle = \langle \bu_{+, 2}^*,
F_{bc}(\tbw )\tbw \rangle.
\end{eqnarray}
\endnumparts
}
 Considering first the generic case $\varkappa \ne 1$, we
substitute Eqs.~(\ref{eq:uFw1})--(\ref{eq:uFw3}) into
Eqs.~(\ref{eq:bichr1})--(\ref{eq:bichr2}) and  find that
Eqs.~(\ref{eq:bichr1}) and (\ref{eq:bichr2}) are compatible only
if $\sin(4\varphi) = 0$, which means that bifurcations of
nonlinear modes are possible only for $\gamma=0$ (i.e. when the
dissipation vanishes) or for $\gamma = \pm 1$. The latter case
corresponds to the point of the phase transition to the broken
$\PT$ symmetry, which is described by the exceptional point
singularity and requires a particular analysis (see
Sec.~\ref{sec:exc}).

The case $\varkappa=1$ corresponds to the peculiar situation when
for any choice of $c_1$ and $c_2$ one has   $F_{bc}(\tbw)\tbw =
\tbw/2$, i.e.   Eqs.~(\ref{eq:bichr1})--(\ref{eq:bichr2})  are
automatically satisfied with $b^{(2)}=1/2$. This is however a
strongly degenerate case. Looking for solutions in the form $w_3^*
= \chi w_1$,  $w_4^* = \chi w_2$ ($\chi$ is arbitrary real)
 %
 %Looking for $\p_{10}\T$ invariant nonlinear modes we set $w_1 = w_3^*$ and $w_2 = w_4^*$. Then
 the system is reduced to the $\PT$-symmetric nonlinear dimer~\cite{Ramezani}
\begin{eqnarray*}
bw_1 = \phantom{+}i\gamma w_1 + w_2 + (\chi^2 +1)|w_1|^2w_1,
\\
bw_2 = -i\gamma w_2 + w_1 + (\chi^2 +1)|w_2|^2w_2,
\end{eqnarray*}
which   supports analytically computable families of  nonlinear modes given by the substitution  $w_1=w_2^*$.
Since $\chi$ is arbitrary,  we can obtain a continuous set of solutions even for fixed $\gamma$ and $b$.

\section{Bifurcations of nonlinear modes in the presence of exceptional-point singularity}
\label{sec:exc}

\subsection{Expansions for nonlinear modes}
\label{sec:ep} Let us turn to a situation when a multiple
eigenvalue $\tb$ corresponds to the so-called exceptional point
singularity which   appears when    coalescence of two (or more)
simple eigenvalues of the Hamiltonian $H(\gamma)$  occurring  at
specific values of the parameter $\gamma$  is accompanied by
collision of the corresponding eigenvectors. Then the geometric
multiplicity of the eigenvalue $\tb$ is less than the algebraic
multiplicity, and the Hamiltonian $H(\gamma)$ becomes
nondiagonalizable. Presence of exceptional points is  a typical
feature of $\PT$-symmetric systems, since such points naturally
appear at the ``boundary'' between phases of broken and unbroken
$\PT$ symmetries~\cite{Graefe11,Ramezani12}.  Nonlinear behavior
of  $\PT$-symmetric optical lattices near the phase-transition
point was recently considered in \cite{Nixon12,Nixon13}.

We  illustrate our ideas  considering an  exceptional point where
\textcolor{black}{two} linear eigenstates coalesce,  forming a
real multiple eigenvalue $\tb$  with  total multiplicity equal to
\textcolor{black}{two and having exactly one linearly independent
eigenvector $\tbw$. We thus have
\begin{eqnarray}
\label{except1}
\left(H- \tb\right)\tbw= 0, \quad \left(H- \tb\right){\bv}= \tbw.
\end{eqnarray}
where we have introduced   a    generalized eigenvector  ${\bv}$.
Since $H$ is $\PT$ symmetric  and $\tbw$ is the only linearly
independent eigenvector corresponding to $\tb$, we can assume that
$\PT\tbw=\tbw$.  We can also assume the the generalized
eigenvector  ${\bv}$ is also   $\PT$ invariant, i.e. $\PT \bv =
\bv$. This assumption  is valid thanks to the following
proposition.}

\medskip
\textcolor{black}{ \textbf{Proposition 2.} There exists a
$\PT$-invariant  generalized eigenvector  ${\bv}$: $\PT \bv =
\bv$.}

\smallskip
\textcolor{black}{ \textbf{Proof of Proposition 2.} Let $\bu$ is
an arbitrarily chosen    generalized eigenvector, i.e.  $\left(H-
\tb\right){\bu}= \tbw$. Applying $\PT$ operator to the both sides
of the latter equality and using that  operator $H-\tb$ commutes
with $\PT$ and that $\PT\tbw=\tbw$, we find that $\PT\bu$ is also
a generalized eigenvector.
 %Therefore, it can be represented as $\PT\bu = \bu + c \tbw$, where $c$ is some constant.  Again, applying $\PT$ operator to the both sides of the latter equation and using that $(\PT)^2=I$, we obtain that $c$ is purely imaginary.
  Then  the $\PT$-invariant generalized eigenvector $\bv$ can be found as $\bv = \frac{1}{2}(\bu + \PT\bu)$.
  %=\bu + \frac{c}{2}\tbw$.
  \hfill $\blacksquare$}
\smallskip

Multiplying the second of equations~(\ref{except1})  by $\p
\tbw=\tbw^*$ and using that $(H-\tb)^\dag = \p(H-\tb)\p$, we find
out that  the   eigenvector $\tbw$  is
self-orthogonal~\cite{Moiseyev} in the sense of the indefinite
$\PT$ inner  product {\cite{BBJ}},
  %$\langle \cdot, \cdot \rangle_{\PT}=\langle \p \cdot, \cdot \rangle$,
  i.e.
  $\langle\p \tbw,\tbw\rangle=0$. Therefore,  now  Eq.~(\ref{eq_p}) as well as the expansions (\ref{expan1})   used in Sec.~\ref{sec:simple} and Sec.~\ref{sec:semisimple}  do not work and have to be modified.
To this end, we look for modified small-amplitude expansions in
the form
\begin{eqnarray}
\label{expan2}
\bw = \varepsilon\tbw +\vep^2\bw^{(2)} +\vep^3\bw^{(3)} + \cdots,\\
\label{expan22}
b = \tb + \varepsilon b^{(1)} + \varepsilon^2b^{(2)}+\cdots.
\end{eqnarray}
Substituting  Eqs.~(\ref{expan2})--(\ref{expan22}) in (\ref{stationary})  and using (\ref{except1}), in the $\epsilon^2$-order we obtain $\bw^{(2)}=b^{(1)}{\bv}$. The third order equation reads
\begin{equation}
\label{second_2}
b^{(2)}\tbw +\left(b^{(1)}\right)^2{\bv}  = (H-\tb)\bw^{(3)} + F(\tbw)\tbw.
\end{equation}
Multiplying   this equation  from the left by $\tbw^*$, we obtain
\begin{eqnarray}
\label{except_cond}
\left(b^{(1)}\right)^2=\frac{\langle \tbw^*,F(\tbw)\tbw\rangle}{\langle \tbw^*,{\bv}\rangle}.
\end{eqnarray}

Let us now assume that the nonlinear operator $F(\bw)$  is of the
$NL_{wPT}$ type [see~(\ref{eq-NL1})]. Then, due to (\ref{prod1}),
we conclude  that the right hand side of (\ref{except_cond}) is
real. Unlike   the case of simple and semi-simple eigenvalues,
however,  for the condition (\ref{except_cond}) to have sense one
has to require its r.h.s. to be nonnegative. This additional
condition,  which does not appear in the case of the ordinary
points, is indeed restrictive, as we will illustrate in
Sec.~\ref{sec:dpT12}. Another important feature  is that
(\ref{except_cond}) indicates that  families  bifurcating from the
eigenvalue  $\tb$  appear in  {\em pairs} (corresponding to two
opposite signs of $b^{(1)}$).

\subsection{Example: Quadrimer with the nearest-neighbor interactions}

\subsubsection{``Phase diagram''}

Let us illustrate the above ideas on the example of a  quadrimer with nearest-neighbor interactions.
The linear part of (\ref{dyn_main}) is now given by the Hamiltonian
 \begin{equation}
 \label{H_quad}
H_{nn} (\gamma_1,\gamma_2)= \left(\begin {array}{cccc}
i\gamma_1 & 1 & 0 & 0\\%
1&i\gamma_2&1&0\\%
0&1 &-i\gamma_2& 1\\%
0&0 &1&-i\gamma_1\end {array} \right),
\end{equation}
where the subscript ``$nn$'' stays for
``\textit{nearest-neibor}''.   The linear operator
$H_{nn}(\gamma_1, \gamma_2)$ is $\p_{11}\T$-symmetric with
$\p_{11}=\sigma_1 \otimes\sigma_1$ [see Eq.~(\ref{eq:P11})], and
the nonlinear part is given by the operator
$F_K(\bw)=\diag(|w_1|^2, |w_2|^2, |w_3|^2, |w_4|^2)$ (where
subscript $K$ stands for Kerr nonlinearity).

\begin{figure}
\includegraphics[width=\textwidth]{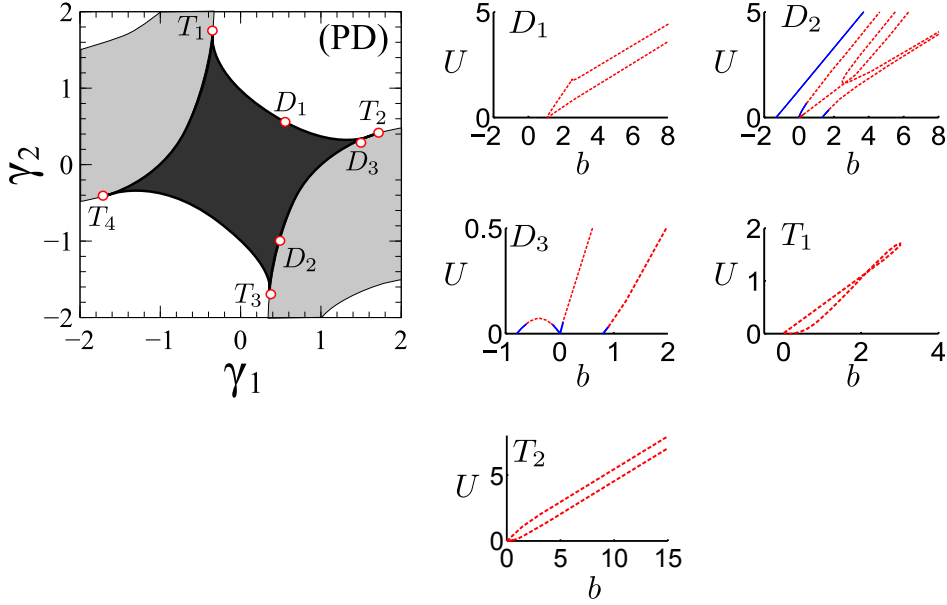}%
\caption{Panel (PD): ``Phase diagram'' of the Hamiltonian
$H_{nn}(\gamma_1, \gamma_2)$ defined by Eq.~(\ref{H_quad}). The
dark-grey diamond-shaped domain corresponds to unbroken $\PT$
symmetry; in the light-gray domains there are two real and two
complex eigenvalues; in the white domains  all eigenvalues are
complex. Other  panels: families of nonlinear modes  on the plane
$U$ \textit{vs.} $b$ for double points $D_{1,2,3}$, and for triple
points \textcolor{black}{$T_1$ and} $T_2$. Stable (unstable) modes
are shown by solid blue (dotted red) lines.}
 \label{fig-PD}
\end{figure}

Properties  of the linear operator  $H_{nn}(\gamma_1,\gamma_2)$
can be visualized conveniently by means of the
``phase-diagram''~\cite{ZK} shown in Fig.~\ref{fig-PD}, panel
(PD).  Depending on $\gamma_1$ and $\gamma_2$, the phase diagram
features three domains: (i) unbroken or \textit{exact} $\PT$
symmetry, when  all the eigenvalues $\tb_j$, $j=1, 2, 3, 4$, of
$H_{nn} (\gamma_1,\gamma_2)$  are real; (ii)  broken $\PT$
symmetry with two real and two complex conjugated eigenvalues
(notice \textcolor{black}{that in the particular case
$\gamma_1=\gamma_2$ domain (ii) can not be encountered)}; (iii)
broken $\PT$ symmetry with all $\tb_j$ complex. Varying parameters
$\gamma_{1,2}$, one can ``travel'' across the phase diagram
visiting domains with different phases. An interesting feature of
the phase diagram shown in Fig.~\ref{fig-PD}~(PD) is that in some
cases increase of the total dissipation brings the system from the
phase of broken $\PT$ symmetry to the unbroken $\PT$ symmetry. For
example, fixing value of one of the coefficients   as $\gamma_1 =
1.2$, than observes that for small $\gamma_2$ (say $\gamma_2=0$)
the $\PT$ symmetry is broken, but when $\gamma_2$ becomes
sufficiently large, then the system enters the phase of the
unbroken $\PT$ symmetry. Further increase of $\gamma_2$  leads to
another phase transition and the $\PT$ symmetry becomes broken
again. A similar scenario with two $\PT$ transitions was recently
reported in \cite{Bender2013}.

 Boundaries separating different domains of the phase diagram correspond to the  exceptional points. In particular, the boundaries contain exactly  four  \emph{triple points} $T_j$,
$j=1,\ldots, 4$,  where the three domains touch. The triple points
correspond to values $\gamma_{1,2}$ for which   $\tb_{1,...,4}=0$,
and the canonical form of  $H_{nn} (\gamma_1,\gamma_2)$ consists
of the only $4\times4$ Jordan block. Depending on  how
$\gamma_{1,2}$ change in the vicinity of $T_j$, either the
$\PT$-symmetric phase or one of the $\PT$ symmetry broken phases
arise.   All the other points of the boundaries are the
\emph{double points} separating two different phases. They are
characterized by the presence of one or two $2\times 2$ Jordan
blocks  in the canonical form of $H_{nn}(\gamma_1,\gamma_2)$. The
double  points can be further sub-classified as belonging to
boundaries separating either phases (i) and (ii), or phases (i)
and (iii), or phases (ii) and (iii). Here however we do not intend
to perform a complete classification and consider only the double
points adjacent to the phase (i) [i.e. the one with unbroken $\PT$
symmetry] since such points are likely to be
\textcolor{black}{more} probable ``candidates'' to give birth to
families of stable nonlinear modes \textcolor{black}{[comparing to
the double points which are not adjacent to the phase (i)]}.

\subsubsection{Existence of nonlinear modes}
The nonlinear modes that obey $\p_{11}\T\bw = \bw$ have the
following property:  $w_1 = w_4^*$, $w_2 = w_3^*$. It is a simple
exercise to ensure that the nonlinear part of the system
$F_K(\bw)$ is of the $NL_{w\PT}(\p_{11})$ type, i.e. obeys
property (\ref{eq-NL1}), see also Sec.~\ref{sec:simple} for
discussion of the properties of $F_K(\bw)$. It was shown~\cite{ZK}
that in this case the nonlinear modes can be found as roots of an
eight-degree polynomial whose coefficients depend on $b$. The
nonlinear modes  constitute  continuous families, which can be
visualized as dependencies $U$ {\it vs.} $b$,  where
$U=\frac{1}{4}\sum_{j=1}^4|w_j|^2$  can be associated with the
norm of the solution (or with  the total energy flow in the
optical context).

\subsubsection{Double points between $T_1$ and $T_2$ (between $T_3$ and $T_4$)}
\label{sec:dpT12}

Let us now consider bifurcations of nonlinear modes from the double points belonging to the boundary separating phases (i) and (iii). In the panel (PD) of Fig.~\ref{fig-PD} such points are situated on the boundary between the points $T_1$ and $T_2$, e.g.   point $D_1$  (or on the boundary between $T_3$ and $T_4$ which is considered analogously). The values of $\gamma_{1,2}$ corresponding to such points are given by the equation
\begin{equation}
(\gamma_1^2 - \gamma_2^2)^2 - 2(\gamma_1^2+4\gamma_1\gamma_2+3\gamma_2^2) + 5 =0.
\end{equation}
In this case the spectrum of the Hamiltonian $H_{nn}(\gamma_1,
\gamma_2)$ consists of two
  opposite   real double eigenvalues $\tb_\pm$, $\tb_+=-\tb_-$. Both $\tb_+$ and $\tb_-$  correspond to a $2\times2$ Jordan block.
Let $\tbw_+$ and $\bv_+$ be the eigenvector and the generalized
eigenvector corresponding to the eigenvalue $\tb_+$ (respectively,
$\tbw_-$ and $\bv_-$ correspond to $\tb_-$).  Then we can write
down two Jordan chains:
\begin{equation}
\label{eq:bpm}%
 (H_{nn} -  \tb_\pm)\tbw_\pm = 0, \quad (H_{nn} -
\tb_\pm)\bv_\pm = \tbw_\pm.
\end{equation}
One can easily check that if one chooses $\p_{11}\T\tbw_\pm =
\tbw_\pm$, then the eigenvectors $\tbw_+$ and $\tbw_-$ can be also
chosen to be related by the following relation:
$\tbw_+=G\p_{11}\tbw_-$, where $G=\diag(-i, i, -i, i)$. Therefore,
the generalized eigenvectors are related through the relation
$\bv_+ = \p_{11} G \bv_-$ [this can be checked by direct
substitution to Eqs.~(\ref{eq:bpm}) and using the relations
$H_{nn}\p_{11} G = G\p_{11} H_{nn}$, $G^2=-I, (\p_{11} G)^2=I$].
Then from Eq.~(\ref{except_cond}) one  obtains
\begin{eqnarray*}
\hspace{-2cm}%
\left(b_+^{(1)}\right)^2 = \frac{\langle
\tbw_+^*,F_K(\tbw_+)\tbw_+\rangle}{\langle \tbw_+^*,
{\bv}_{+}\rangle} =  \frac{\langle -G\p_{11}
\tbw_-^*,F_K(\tbw_-)G\p_{11} \tbw_-\rangle}{\langle -G\p_{11}
\tbw_-^*, \p_{11} G {\bv}_{-}\rangle}=\frac{\langle
\tbw_-^*,F_K(\tbw_-)\tbw_-\rangle}{\langle \p_{11} G G \p_{11}
\tbw_-^*,{\bv}_{-}\rangle} \\= -\frac{\langle
\tbw_-^*,F_K(\tbw_-)\tbw_-\rangle}{\langle
\tbw_-^*,{\bv}_{-}\rangle} = -\left(b_-^{(1)}\right)^2.
\end{eqnarray*}
Here we have additionally used that $F_K(\tbw_+)=F_K(\tbw_-)$,
$(G\p_{11})^2=I$, $\p_{11}^2=I$,  and that the operators  $\p_{11}
G$ and $G\p_{11}$ are Hermitian. The obtained result indicates
that if $b_+^{(1)}\ne 0$, then the bifurcation of nonlinear modes
is possible only from one eigenstate, either from $\tb_+$ or from
$\tb_-=-\tb_+$.

As an  example, in Fig.~\ref{fig-PD} we report numerically
obtained families of  nonlinear modes at the double point $D_1$
where $\gamma_{1}=\gamma_{2}=\sqrt{5}/4$. Two families of
unstable nonlinear modes bifurcate from the positive eigenvalue
and no nonlinear modes bifurcates from the negative one (hereafter
the stability of nonlinear modes have been investigated by means
of analysis of the spectrum of the linearized problem).

\subsubsection{Double points between $T_2$ and $T_3$ (between $T_1$ and $T_4$).}
Such double points lie on   the hyperbola $\gamma_2 = 1 -
1/\gamma_1$ (for the double points  located between the triple
points $T_2$ and $T_3$), or on $\gamma_2 = -1 - 1/\gamma_1$ (for
the double points  between  $T_1$ and $T_4$). In this case
spectrum of $H_{nn}(\gamma_1, \gamma_2)$ contains  a double zero
eigenvalue   $\tb_0=0$ corresponding to a  $2\times 2$ Jordan
block, \textcolor{black}{ and hence
\begin{eqnarray}
\label{det=0}%
 \det H_{nn}(\gamma_1, \gamma_2)=0.
\end{eqnarray}
}
Two other eigenvalues of $H_{nn}(\gamma_1, \gamma_2)$ are simple,
real and have    opposite values.

The peculiarity of these double points is that
\textcolor{black}{due to  the particular structure of the linear
eigenvector $\tbw$ and the nonlinearity $F_K(\tbw)$},
Eq.~(\ref{except_cond}) yields that the coefficient   $b^{(1)}$ is
equal to zero. \textcolor{black}{Then for any $b^{(2)}$
Eq.~(\ref{second_2}) has a nontrivial solution $\bw^{(3)}$. This
also implies that $\bw^{(2)}=0$.} Therefore, for a more detail
description of the families bifurcating from the double eigenvalue
$\tb_0$ \textcolor{black}{one has to proceed to next orders of the
expansions (\ref{expan2})--(\ref{expan22}). The modified
expansions read
\begin{eqnarray}
\label{expan20}
\bw = \varepsilon\tbw +\vep^3\bw^{(3)} +\vep^4\bw^{(4)} + \vep^5\bw^{(5)}\cdots,\\ %
b = \varepsilon^2b^{(2)}+\varepsilon^3b^{(3)}+\varepsilon^4b^{(4)}\cdots,
\end{eqnarray}
which in the $\varepsilon^4$-order gives $\bw^{(4)}=b^{(3)}\bv_1$, while in the $\varepsilon^5$-order we obtain
\begin{equation}
\label{eq:eps5} b^{(4)}\tbw + b^{(2)}\bw^{3} = H_{nn}\tbw^{(5)} +
{\bf f}_5,
\end{equation}
where ${\bf f_5}$ is the  $\varepsilon^5$-order contribution of
the nonlinear term $F_K(\bw)\bw$:
\begin{equation}
{\bf f_5}=\tbw\circ\tbw\circ\left(\bw^{(3)}\right)^* + 2\tbw\circ\tbw^*\circ\bw^{(3)},
\end{equation}
where we used ``$\circ$'' to designate element-wise multiplication
of vectors. From Eq.~(\ref{eq:eps5}) we obtain
\begin{equation}
\label{eq:b25}
b^{(2)}{\langle \tbw^*, \bw^{(3)}\rangle} = {\langle \tbw^*, {\bf f}_5\rangle}.
\end{equation}
This  equation together with Eq.~(\ref{second_2}) can be used to
compute $b^{(2)}$. To this end, let us first solve
Eq.~(\ref{second_2}) with $b^{(2)}=0$,  i.e.
\begin{equation}
\label{eq:Hw3}
 H_{nn}\bw^{(3)} =- F_K(\tbw)\tbw
\end{equation}
 with respect to $\bw^{(3)}$.
} \textcolor{black}{ In spite of the equality (\ref{det=0}),
Eq.~(\ref{eq:Hw3}) does have a $\PT$-invariant solution.}
\textcolor{black}{Indeed, if $\bu$ is an arbitrary solution of
Eq.~(\ref{eq:Hw3}), then applying $\PT$ operator two both sides of
(\ref{eq:Hw3}) and using the fact that the nonlinearity $F_K(\bw)$
is weakly $\PT$ symmetric [i.e. obeys (\ref{eq-NL1})], we find
that  $\PT\bu$ is also solution of the same equation. Thus a
$\PT$-invariant solution of  Eq.~(\ref{eq:Hw3}) is given by
$\bu_0=\frac{1}{2}(\bu + \PT\bu)$.}

On the other hand, \textcolor{black}{to construct a}
\textcolor{black}{solution of Eq.~(\ref{eq:Hw3}) we let the first
entry of $\bw^{(3)}$ to be a free parameter and express all other
entries of $\bw^{(3)}$ in terms of the first one. Introducing $c =
w_1^{(3)}$ (where $c$ is the free parameter and   $w_1^{(3)}$ is
the first entry of $\bw^{(3)}$), for the next entries we find
$w_2^{(3)} = -i\gamma_1 c - |\tw_1|^2\tw_1$, and   $w_3^{(3)} =
i\gamma_2|\tw_1|^2\tw_1 - \gamma_1 c - |\tw_{2}|^2\tw_2$, where
$\tw_j$ are the entries of the linear eigenvector  $\tbw$. (Here
we have also  used relation $\gamma_1\gamma_2+1=\gamma_1$ which is
valid for the double points on the boundary between  $T_2$ and
$T_3$; for the double points situated on the  boundary  between
$T_1$ and $T_4$ the resulting expressions will be slightly
different but having the same structure.) Looking for a
$\PT$-invariant solution (which is shown to   exist) we require
$\left(w_3^{(3)}\right)^* = w_2^{(3)}$, which results in   a
system of two linear equations with respect to $c_1 = \RE  c$ and
$c_2 = \IM  c$:
\begin{eqnarray}
\gamma_1(c_1 + c_2) =  \RE\left (|\tw_1|^2\tw_1 - |\tw_2|^2\tw_2^* - i\gamma_2 |\tw_1|^2\tw_1^* \right),\\
\gamma_1(c_1 + c_2)  = \IM\left (|\tw_1|^2\tw_1 - |\tw_2|^2\tw_2^* - i\gamma_2 |\tw_1|^2\tw_1^* \right).
\end{eqnarray}
 While the determinant of latter system is zero, the system has to be compatible (otherwise, the $\PT$-invariant solution would not exist). We   set $c_2=0$, and  then the parameter $c$ is fixed as follows:
\begin{equation}
\label{eq:c}
c=  \RE  c  = \frac{1}{\gamma_1}  \RE\left (|\tw_1|^2\tw_1 - |\tw_2|^2(\tw_2)^* - i\gamma_2 |\tw_1|^2\tw_1^* \right).
\end{equation}
Therefore, the $\PT$-invariant solution for Eq.~(\ref{eq:Hw3}) is written down as
\begin{eqnarray}
\label{eq:u0}
 \bu_{0}=\left(\!\!
 \begin{array}{c}
c \\ -i\gamma_1 c - |\tw_1|^2\tw_1\\ \phantom{-}i\gamma_1 c - |\tw_1|^2\tw_1^*\\ c
 \end{array}
 \!\!\right).
\end{eqnarray}
Notice that choice $c_2=0$ was not restrictive since the most general $\PT$-invariant solution of Eq.~(\ref{eq:Hw3}) can be found   as $\bw^{(3)}=\bu_0 + d \tbw$,  where  the particular solution  $\bu_0$ is fixed by Eqs.~(\ref{eq:c})--(\ref{eq:u0}), and  where $d$ is arbitrary real number, which  can always be set equal to zero by means of rescaling of the small parameter $\varepsilon$.
}

\textcolor{black}{ Solution of  Eq.~(\ref{second_2})  with
arbitrary $b^{(2)}$ can be found as $\bw^{(3)} = \bu_0 +
b^{(2)}\bv$. Substituting the latter expression into
Eq.~(\ref{eq:b25}), one obtains a quadratic equation with respect
to $b^{(2)}$ (notice that all the   coefficients of the latter
equation  are real and expressed in terms of $\tbw$, $\bv$ and
$\gamma_{1,2}$).}

\textcolor{black}{ Resorting  at this stage to numerics,  we
consider two  double points: $D_2 = (\frac{1}{2}, -1)$ and  $D_3 =
(\frac{3}{2}, \frac{1}{3})$. Finding for  both of them the
quadratic equation for $b^{(2)}$ and computing its roots, we
observe that both for  $D_2$ and  $D_3$ the roots are real,
nonzero, and distinct from each other. Moreover, for the point
$D_2$ both roots are of the same sign, while for $D_3$ the roots
have opposite signs. Therefore, both for $D_2$  and $D_3$  two
families of nonlinear modes are expected to bifurcate from the
double eigenvalue $\tb=0$.  This is confirmed by numerical results
in  Fig.~\ref{fig-PD}  where we show   families of nonlinear modes
both    for the points $D_2$  and  $D_3$. Bifurcation diagrams for
$D_2$ and $D_3$  are similar locally near the double eigenvalue
$\tb = 0$ in terms of existence: both for $D_2$ and $D_3$ there
are two families bifurcating from $\tb$. Notice however that for
the point  $D_2$ both the families bifurcating from $\tb$
bifurcate to the right (i.e. to the half-plane $b>0$), while at
the point $D_3$ one of the families bifurcates to the right and
another one bifurcates to the left (i.e. to the half-plane $b<0$).
This behavior  is in agreement with different signs  of the roots
of the quadratic equations for  $b^{(2)}$. We also notice  that
stability of the bifurcating families is different: for the point
$D_2$ one of the families is stable (in the vicinity of the
bifurcation), while another one is unstable; on the other hand,
for  the point $D_3$ both the bifurcating families are stable in
the sufficiently small vicinity of the bifurcation. Existence  of
stable nonlinear modes in spite of the presence of the exceptional
point singularity is quite  remarkable.   }

 %Notice also that  points $D_2$ and $D_3$   feature  different behavior of the families   of nonlinear modes in the domain of finite nonlinearity.

\subsubsection{Triple points} There exist  exactly four \textit{triple points}  $T_j$, $j=1,2,3,4$. Each triple point corresponds to the situation when the   Hamiltonian $H_{nn}(\gamma_1, \gamma_2)$ has a zero eigenvalue $\tb_0=0$ with multiplicity equal to 4.  The canonical Jordan representation of   $H_{nn}(\gamma_1, \gamma_2)$ in the triple points is given by a   $4\times 4$ Jordan block. \textcolor{black}{Therefore, our analysis in Eqs.~(\ref{except1})--(\ref{except_cond}) is not applicable in this case.  However, for the sake of   completeness of our studies, we report   numerical results on  the behavior of the nonlinear modes bifurcating from the triple
points.}

The points $T_2$ and $T_3$ correspond to values of $\gamma_1$
given as   real roots of the equation
$\gamma_1^4-2\gamma_1^2-2\gamma_1+1=0$, and $\gamma_2 =
1-1/\gamma_1$. For the triple points $T_{1}$ and $T_{4}$ one has
the equation $\gamma_1^4-2\gamma_1^2+2\gamma_1+1=0$, and
$\gamma_2=-1-1/\gamma_1$. \textcolor{black}{Notice that for each
triple points $T_j$  one has $\gamma_1^2+\gamma_2^2=3$.}

\textcolor{black}{In Fig.~\ref{fig-PD} we show numerical results
for   points  $T_1\approx (-0.37, 1.69)$ and  $T_2\approx (1.68,
0.41)$ which feature different bifurcation diagrams. All the found
modes are unstable.}

\section{Nonlinearities allowing for integrals of motion}
\label{sec:intergral}

\subsection{General idea}

It is  known that the integrals of motion are of fundamental
importance for the conservative systems, while they do not
necessarily  exist for the dissipative ones. Although the
dissipative systems typically do not conserve the total energy,
they can admit other conserved quantities and even be integrable
\cite{Akhmediev}. For linear $\PT$-symmetric systems, some
integrals of motion can be found in an explicit form
\cite{znojil}, but  in the nonlinear case the only system with
known integrals of motion  (to the best of the authors' knowledge)
is the  exactly integrable dimer~\cite{Ramezani}
\textcolor{black}{(see also~\cite{archive} where some of the
results stemming from the integrable dynamics of a dimer were
generalized to the chain of coupled dimers).}

So far,  we  considered how stationary  nonlinear modes depend on
the character of the linear eigenstate they bifurcate from and on
the type of the nonlinearity  $F(\bq)$. Let us now consider how
the  choice of nonlinearity $F(\bq)$ can affect  dynamical
properties of the system. More specifically, we address the
existence of the conserved quantities  of the nonlinear  system
(\ref{dyn_main}).
%
%To the best of the authors' knowledge, so far there has been
%reported only one example of a $\PT$-symmetric system obeying
%explicit integrals of motions --- the integrable nonlinear
%dimer~\cite{Ramezani} \textcolor{black}{(see also~\cite{archive}
%where some of the results stemming form the integrable dynamics of
%a dimer were generalized to the chain of coupled dimers)}.
In this
section, we find a condition which must be satisfied by the
nonlinear operator $F(\bq)$ for the nonlinear system
(\ref{dyn_main}) to support at least one integral of motion. Using
this result, we  report several integrals of motion for a
$\PT$-symmetric quadrimer.

A motivation for our consideration is the known fact that a linear
$\PT$  system [which formally corresponds to Eq.~(\ref{dyn_main})
with $F(\bq)=0$] has an integral
which is given as the ``pseudo-power''  $Q  = \langle \p \bq(t), \bq(t)\rangle$~\cite{znojil} [this  fact can be  also easily verified using Eqs.~(\ref{PT_sym})]. %(see e.g.~\cite{} and \cite{ZK} for the discrete case).
However, the equality $ \dot{Q} = 0$  generically  does not hold
for the nonlinear  system (\ref{dyn_main}) with $F(\bq)\ne0$. To
establish conditions for the nonlinear system to admit an integral
of motion , we look for a conserved quantity  in the form $Q=
\langle A \bq(t), \bq(t)\rangle$ where $A$ is  an arbitrary (so
far) time-independent linear  operator. Then (\ref{dyn_main})
yields
\begin{equation}
i\dot{Q} = -\langle A[H+F(\bq)]\bq, \bq\rangle  +   \langle [H^\dag+F^\dag(\bq)] A\bq, \bq\rangle. %\mbox{ \textcolor{blue}{check sign here}}
\end{equation}
For $Q$ to  be a conserved quantity it is sufficient to require
%\numparts
\begin{eqnarray}
\label{eq:AH1}
AH = H^\dag A
\end{eqnarray}
and
\begin{eqnarray}
\label{eq:AH2}
AF(\bq) = F^\dag (\bq) A \mbox{\quad for all } \bq.
\end{eqnarray}
%\endnumparts

A particularly interesting case corresponds to a  situation when
the properties (\ref{eq:AH1})--(\ref{eq:AH2}) hold for $A=\p$,
where $\p$ is a   parity operator. Then   condition (\ref{eq:AH1})
is equivalent to the pseudo-Hermiticity~\cite{Mostaf2002} of $H$,
see also Eq.~(\ref{PT_sym}). Equation (\ref{eq:AH2}) now gives
\begin{eqnarray}
\label{NL_PH}
F^\dag(\bq) = \p F(\bq) \p \mbox{\quad for all } \bq,
\end{eqnarray}
i.e. \emph{for all $\bq$} the nonlinear operator $F(\bq)$ must be
pseudo-Hermitian with respect to the same $\p$ operator as the
linear operator $H$.
The nonlinearities $F(\bq)$  obeying   to  the property %(\ref{eq-NL3}) or
(\ref{NL_PH}) will be said to be of the pseudo-Hermitian type and
the class of such nonlinearities will be denoted as $NL_{pH}(\p)$
(with  the respective parity operator $\p$ indicated in the
brackets).

The following proposition establishes a simple relation    between
pseudo-Hermitian and $\PT$-symmetric nonlinearities introduced in
Sec.~\ref{sec:simple}.

\medskip

\textbf{Proposition 3.} If $F(\bq)\in NL_{\PT}(\p)$ and
$F(\bq)=F^T(\bq)$, then $F\in NL_{pH}(\p)$.

\subsection{Example:   $\PT$-symmetric quadrimer with integrals of motion}
\label{integralBF}

Turning now to particular examples, we consider dynamics of  the
system (\ref{dyn_main}) described by the linear Hamiltonian
$H_{bc}(\gamma_1, \gamma_2)$ defined by Eq.~(\ref{Hbc}) and by the
nonlinearity  $F_{bc}(\bq)$ given by (\ref{F_polychrom}). As it
has already been noticed in Sec.~\ref{sec:polychrom}, the linear
operator $H_{bc}(\gamma_1,\gamma_2)$ is $\p_{10}\T$-symmetric [see
Eq.~(\ref{eq:Ps}) for the  definition of   $\p_{10}$], and the
nonlinearity  $F_{bc}(\bq)$ is $\PT$ symmetric, i.e.
$F_{bc}(\bq)\in NL_{\PT}(\p_{10})$. Due to Proposition~3, the
latter fact \textcolor{black}{together with the diagonal structure
of $F_{bc}(\bq)$}  implies that $F_{bc}(\bq) \in NL_{pH}(\p_{10})$
(this can also be checked in a straightforward manner). Hence
there exists at least one integral of motion  given by
\begin{equation}
Q_1 = \langle \p_{10}\bq, \bq\rangle = 2\RE(q_1^*q_3 + q_2^*q_4).
\end{equation}

Furthermore, the model %(\ref{H_BF}),
(\ref{Hbc})--(\ref{F_polychrom}) possesses another integral of
motion. To find it,  we notice that operator
$H_{bc}(\gamma_1,\gamma_2)$ is also $\p_{20}\T$-symmetric with
respect to
\begin{equation}
\p_{20}=\sigma_2\otimes\sigma_0= \left(\begin {array}{cccc}
0 & 0 & -i &0\\%
0&0&0&-i\\%
i&0 &0& 0\\%
0&i &0&0\end {array} \right).
\end{equation}
Notice that while the operator $\p_{20}$ is Hermitian, it can not
be considered as a conventional parity operator  because  it does
not commute with the operator $\T$ which  in our case is  the
complex conjugation. However, one can check that $F_{bc}^\dag(\bw)
= F_{bc}(\bw) = \p_{20} F_{bc}(\bw) \p_{20}$, i.e.  $F_{bc} (\bq)
\in NL_{pH}(\p_{20})$. Thus the second integral of motion is
readily found to be
\begin{equation}
Q_2 = \langle \p_{20}\bq, \bq\rangle = 2\IM(q_1^*q_3 + q_2^*q_4).
\end{equation}
Obviously, $Q_1$ and $Q_2$ can be combined in a single complex-valued integral
\begin{equation}
\label{Q}
Q = q_1^*q_3 + q_2^*q_4.
\end{equation}

The integral (\ref{Q}) can be also obtained from a more general
consideration. Indeed, considering  the operator
\begin{equation}
A = \left(\begin {array}{cccc}
0 & 0 & a_1 &0\\%
0&0&0&a_1\\%
a_2&0 &0& 0\\%
0&a_2 &0&0\end {array} \right)
\end{equation}
where $a_{1,2}$ are  {arbitrary }, one can verify that the both conditions (\ref{eq:AH1})--(\ref{eq:AH2}) hold. Then a conserved quantity is given as
\begin{equation}
\langle A\bq, \bq\rangle = a_1 (q_1^*q_3 + q_2^*q_4) + a_2 (q_1q_3^* + q_2q_4^*).
\end{equation}
Setting $a_2=0$ one readily obtains   integral (\ref{Q}).

It turns out, however, that  in a particular  case when the linear
part is given by $H_{bc}(\gamma,-\gamma)$, and $\varkappa=1$ in
Eq.~(\ref{F_polychrom}),  the obtained quantity $Q$ is not the
only integral  of motion. Indeed, let us consider the model
(\ref{ex-1})--(\ref{ex-4}) with $\varkappa=1$:%
\numparts
\begin{eqnarray}
\label{BF-1}
i\dot{q}_1=i\gamma q_1+ q_2+(|q_1|^2 +|q_3|^2)q_1,
\\
\label{BF-2}
i\dot{q}_2=-i\gamma q_2+ q_1+(|q_2|^2 +|q_4|^2)q_2,
\\
\label{BF-3}
i\dot{q}_3=-i\gamma q_3+ q_4+(|q_3|^2 +|q_1|^2)q_3,
\\
\label{BF-4}
i\dot{q}_4=i\gamma q_4+ q_3+(|q_4|^2 +|q_2|^2)q_4.
\end{eqnarray}
\endnumparts
%which is defined by the Hamiltonian   given by (\ref{H_BF}).
To construct a new integral, we   rewrite Eqs.~(\ref{BF-1})-(\ref{BF-4}) in terms of the Stokes components (this approach was used in~\cite{Ramezani} in the study of the nonlinear  $\PT$-symmetric dimer):
\begin{eqnarray*}
S_{1}^{0}=|q_1|^2+|q_2|^2, \quad        &   S_{2}^{0}=|q_3|^2+|q_4|^2,\\%
S_{1}^{1}=q_1^*q_2+q_1q_2^* \quad       &   S_{2}^{1}=q_4^*q_3+q_4q_3^*,\\%
S_{1}^{2}=i(q_1q_2^*-q_1^*q_2), \quad   &   S_{2}^{2}=i(q_4q_3^*-q_4^*q_3),\\%
S_{1}^{3}=|q_1|^2-|q_2|^2, \quad        &   S_{2}^{3}=|q_4|^2-|q_3|^2.
%
%S_{2}^{0}=|q_3|^2+|q_3|^2, \quad
%S_{2}^{2}=i(q_4q_3^*-q_4^*q_3), \quad
\end{eqnarray*}
We also introduce the Stokes vectors $\bS_j=(S_j^1,S_j^2,S_j^3)$ with $j=1,2$,  and notice that % scalars
\begin{eqnarray}
\label{S0S}
%\| \bS_j\|^2 \equiv
\bS_j \cdot \bS_j = (S_j^0)^2.
\end{eqnarray}
Next, we introduce   the pseudo-electric field $\bE=(0,0,2\gamma)$, pseudo-magnetic fields $\bB_0=(-2,0,0)$ and $\bB=(0,0,S_2^3-S_1^3)$, and rewrite the system (\ref{BF-1})-(\ref{BF-4}) in the form
\begin{eqnarray}
\label{S0}
\dot{S}_1^0=\bE\cdot\bS_1, \quad \dot{S}_2^0=\bE\cdot\bS_2,
\\
\label{SS}
\dot{\bS}_1=S_1^0\bE+\bS_1\times(\bB_0+\bB),\quad \dot{\bS}_2=S_2^0\bE+\bS_2\times(\bB_0-\bB).
\end{eqnarray}
Further,  multiplying Eqs.~(\ref{SS}) by $\bB$, we obtain
($j=1,2$)
\begin{eqnarray}
\label{b1}
\bB \cdot \dot{\bS}_j=S_j^0\bB\cdot\bE+\bS_j\cdot (\bB_0\times\bB),
\end{eqnarray}
and   multiplying  Eqs.~(\ref{SS})  by $\bB_0$ we obtain
\begin{eqnarray}
\label{b2}
\bB_0 \cdot \dot{\bS}_1=-\bS_1\cdot (\bB_0\times \bB), \qquad  \bB_0\cdot \dot{\bS}_2=\bS_2\cdot (\bB_0\times \bB).
\end{eqnarray}
 Combining (\ref{b1}) and (\ref{b2}) yields the relations
\begin{eqnarray}
\label{eq:inttmp}
\bB\cdot\dot{\bS}_1=S_1^0\,\bB\cdot\bE-\bB_0\cdot \dot{\bS}_1,\quad \bB\cdot \dot{\bS}_2=S_2^0\,\bB\cdot\bE+\bB_0\cdot \dot{\bS}_2.
\end{eqnarray}
After subtracting one of Eqs.~(\ref{eq:inttmp})  from another and using the definitions of the vectors $\bB$, $\bB_0$, and $\bE$, we arrive at
\begin{eqnarray*}
0=-\bB\cdot (\dot{\bS}_1-\dot{\bS}_2)+(S_1^0-S_2^0)\,\bB\cdot\bE-\bB_0\cdot (\dot{\bS}_1+\dot{\bS}_2)
\\
=(S_2^3-S_1^3)\frac{d}{dt}(S_2^3-S_1^3) - (S_1^0-S_2^0)\frac{d}{dt}(S_1^0-S_2^0)+2\frac{d}{dt}(S_1^1+S_2^1)
\\
=\dot{J},
\end{eqnarray*}
where   integral of motion $J$ is given by
\begin{eqnarray}
\label{int_I}
J=\frac 12(S_2^3-S_1^3)^2-\frac 12(S_2^0-S_1^0)^2+2(S_2^1+S_1^1).
%\nonumber \\
%= -\frac 12 [(S_2^1)^2+(S_2^2)^2+(S_1^1)^2+(S_2^2)^2]+2(S_1^1+S_2^2)
%\nonumber \\
%%-S_2^3S_1^3-[\bS_1^2\bS_2^2]^{1/2}
%-S_2^3S_1^3- \|\bS_1\|\, \|\bS_2\|,
\end{eqnarray}
%(in order to obtain the last equality we used (\ref{S0S})  to express the integral only through the components of the three-dimensional Stokes vectors $\bS_{1,2}$).
Using  Eqs.~(\ref{S0S}) one can express $J$ only through the components of the three-dimensional Stokes vectors $\bS_{1,2}$.

Notice that in the limit $q_3=q_4=0$, i.e. $S_2^0=0$ and  $\bS_2=0$,
the obtained integral is reduced to one of the known integrals for a $\PT$-symmetric dimer~\cite{Ramezani}.

Finally, we notice that  system (\ref{BF-1})--(\ref{BF-4}) admits a solution in quadratures  in the particular case when $Q=0$. Indeed, this implies that $|q_1|^2|q_3|^2=|q_2|^2|q_4|^2$, which is equivalent to $S_1^0S_2^3=S_1^3S_2^0$. Combing the latter relation with Eqs.~(\ref{S0}), we have $S_1^0\dot{S}_2^0=S_2^0\dot{S}_1^0$, and hence $S_1^0=CS_2^0$, where $C$ is a constant. Subsequently $\bE\cdot \bS_1=C\bE\cdot \bS_2$, i.e. $S_1^3=CS_2^3$ and $\bB=(1-C)(0,0,S_2^3)$. The  latter formula means that the dynamical equation for $\bS_2$ is singled out and acquires the form of the equation for the dimer. Solutions for such a dimer in an implicit form have been found in~\cite{Ramezani}.

%A potential extension of the above system is
%\begin{equation*}
%H = \left(\begin {array}{cccc}
%i\gamma_1 & 1 & 0 & \vep\\%
%1&i\gamma_2&\vep&0\\%
%0&\vep &-i\gamma_1& 1\\%
%\vep&0 &1&-i\gamma_2\end {array} \right)
%\end{equation*}
%It has the same properties, including the integral. Actually this is the same system that was suggested in the old draft of Panos et al.

\section{Conclusion}

In the presented paper we have investigated some of  nonlinear
properties of finite-dimensional  systems respecting $\PT$
symmetry. In contrast to the most part of the previous studies,
our discussion has not been focused on a nonlinearity of any given
form. Instead, we have emphasized how  nonlinearities of different
classes can affect stationary and dynamical properties of the
system. First, we have considered a class  of nonlinearities with
weak $\PT$ symmetry, the latter appearing to be necessary (and in
many cases sufficient)  for the existence of the families of
stationary nonlinear modes. We have paid particular attention to
analysis of the bifurcations of nonlinear modes from the multiple
eigenstates of the underlying linear problem. In a situation when
the   underlying linear system has a semi-simple eigenstate, we
have shown that the invariant subspace associated with  the
degenerate eigenvalue can always be  spanned by a complete basis
of $\PT$-invariant eigenvectors. The established fact has been
used to construct formal expansions for the nonlinear modes
bifurcating from the degenerate eigenstate. Next, we considered
bifurcations of stationary modes from  exceptional points, which
typically occur at the phase transition between unbroken and
broken $\PT$ symmetries.  We have shown that the generalized
eigenvector associated with the multiple eigenvalue can be chosen
$\PT$ invariant.  Then we have developed small-amplitude
expansions for the bifurcations of nonlinear modes and
demonstrated that the possibility of  the bifurcations  as well as
stability of the modes depend on both the nonlinearity and the
character of the coalescing eigenstates.

To complete the  above picture, we notice that the nonlinear modes
can also exist in the case when $\PT$ symmetry of the underlying
linear problem is broken, i.e. the   spectrum  of the underlying
linear operator  $H(\gamma)$ contains complex eigenvalues. If all
the linear eigenvalues are complex, then the nonlinear modes
obviously cannot bifurcate from the linear eigenstates, i.e. such
nonlinear modes (if exist) do not have linear counterpart.
Moreover, in the case of a finite-dimensional system (like a
$\PT$-symmetric oligomer) nonlinear modes can be stable even if
the $\PT$ symmetry of the linear problem is broken~\cite{ZK}.

%families of nonlinear modes
%bifurcating from the simple (non-degenerate) eigenstate, from the semi-simple (degenerate) eigenstates, and from the %exceptional points  of the underlying linear system.

Next, we have turned to dynamical  properties of nonlinear
$\PT$-symmetric lattices and  indicated another important class of
nonlinearities, termed pseudo-Hermitian nonlinearities, which
allow for the nonlinear  system to admit  at least one integral of
motion. Using this idea, we have  found several integrals for  a
$\PT$-symmetric nonlinear quadrimer and demonstrated that  (at
least in some cases) it admits  a solution in quadratures.

\section*{Acknowledgements}

Authors are grateful  to P. G. Kevrekidis and U. G\"unther for
stimulating discussion. The work was supported by FCT  (Portugal)
through the FCT  grants PEst-OE/FIS/UI0618/2011,
PTDC/FIS-OPT/1918/2012, and SFRH/BPD/64835/2009.

\end{document}